# POISSON SOLVERS FOR SELF-CONSISTENT MULTI-PARTICLE SIMULATIONS[*]

J. Qiang, S. Paret, LBNL, Berkeley, USA

*Abstract*

Self-consistent multi-particle simulation plays an important role in studying beam–beam effects and space charge effects in high-intensity beams. The Poisson equation has to be solved at each time-step based on the particle density distribution in the multi-particle simulation. In this paper, we review a number of numerical methods that can be used to solve the Poisson equation efficiently. The computational complexity of those numerical methods will be $O(N \log(N))$ or $O(N)$ instead of $O(N^2)$, where $N$ is the total number of grid points used to solve the Poisson equation.

## INTRODUCTION

The self-consistent multi-particle simulation based on the particle-in-cell method has been widely used in studying beam–beam effects in high-energy ring colliders and space charge effects in high-intensity/brightness accelerators. A schematic plot of a single step of the multi-particle simulation for the strong–strong (self-consistent) beam–beam simulation is given in Fig. 1.

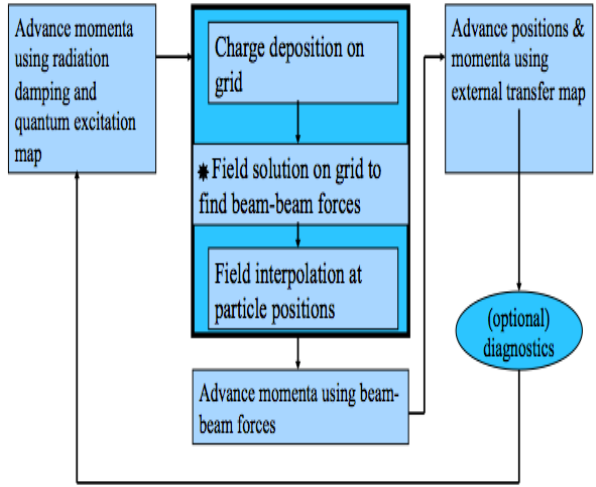

Figure 1: A schematic plot of a single step in the multi-particle simulation.

Here, we assume that a number of macro-particles have been initially generated from a given initial 6D phase space distribution. Within this step, these macro-particles are deposited onto a 2D computational grid for each slice to obtain the spatial charge density distribution. The Poisson equation is solved on the grid and the electromagnetic fields are calculated from the solution of the Poisson equation. Those fields are interpolated back to individual particle positions to calculate the beam–beam forces. Momenta of each particle are updated using the beam–beam forces. Positions and momenta of each particle are advanced using a transfer map. For a lepton accelerator, particle momenta are further advanced using a radiation damping and quantum excitation map to finish the single step. This single-step loop is repeated many times in the multi-particle simulation. Since the charge density will be updated during every step, the Poisson equation has to be solved for every step. The speed of the Poisson solver could become the bottleneck for the whole simulation. In the following section we review a number of efficient numerical methods (FFT-based method, spectral-finite difference method, and multigrid spectral-finite difference method) used to solve the Poisson equation in multi-particle beam–beam and space charge simulations.

## FFT-BASED GREEN FUNCTION METHOD

The FFT-based Green function method is mostly used to solve the Poisson equation subject to an open boundary condition. This is true if the pipe radius in an accelerator is much larger than the beam bunch transverse size. Given the Poisson equation:

$$\nabla^2 \phi = -\frac{\rho}{\varepsilon} \qquad (1)$$

subject to the open boundary conditions, the solution of the electric potential can be written as:

$$\phi(r) = \int G(r-r')\rho(r')dr' \qquad (2)$$

where the Green function is given by:

$$G(r) = \begin{cases} -\log(r): & \text{in } 2D \\ \dfrac{1}{r}: & \text{in } 3D \end{cases} \qquad (3)$$

Integral Eq. (2) can be written as a numerical summation for each grid point in a 2D problem:

---

[*] Work supported by the US Department of Energy under contract No. DE-AC02- 05CH11231.

$$\phi(x_i,y_j)=h_x h_y\sum_{i'=1}^{N_x}\sum_{j'=1}^{N_y}G(x_i-x_{i'},y_j-y_{j'})\rho(x_{i'},y_{j'}) \quad (4)$$

A direct calculation of the above summation will have a computational cost of $O(N^2)$, where $N = N_x N_y$. For a grid of $100 \times 100$ points in each dimension, this yields an operation of $10^8$ in a 2D problem and $10^{12}$ in a 3D problem to solve the Poisson equation for each step.

The direct calculation is very inefficient and will significantly slow down the computational speed of the multi-particle simulation. Fortunately, Eq. (4) can be calculated using an FFT-based method by turning it into a cyclic summation. The idea behind this method is to construct a discrete periodic system so that the FFT can be used to calculate this discrete cyclic summation. In this new periodic system, the original computational domain is doubled in each dimension within a period of computational domain. A new Green function and a new charge density function are defined in this new computational domain so that the cyclic summation will yield the same results as the original summation in Eq. (4) inside the original computational domain. Outside the original computational domain, the two summation results will be different but irrelevant since we are only interested in the field inside the original domain. The detailed expression for the Green's function and the charge density inside the new computational domain can be found in Refs. [1, 2].

The computational cost for the cyclic summation will be $O(N \log(N))$. Figure 2 shows a comparison of the electric field as a function of the radius from the FFT-based numerical solution and from the analytical solution of a 2D Gaussian density distribution. The agreement between those two solutions is excellent.

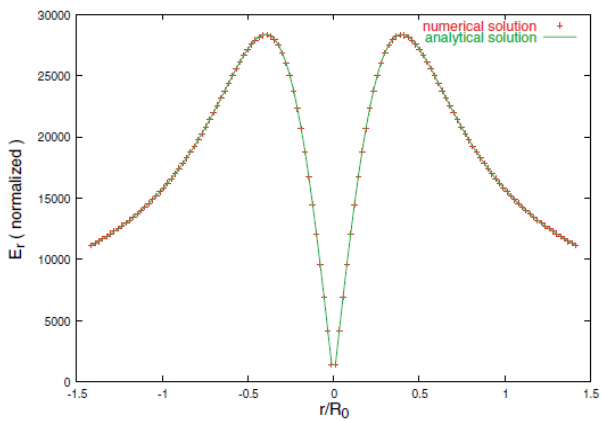

Figure 2: Electric field as a function radius within a 2D Gaussian distribution from the numerical solution (red) and the analytical solution (green).

The FFT-based method described above works well for a beam with a small aspect ratio. For a beam with a large aspect ratio, which is the case for an electron or positron beam inside a lepton collider, the direct use of the above FFT-based method will be inefficient since a large number of grid points are needed to resolve the variation of the Green function inside a grid cell. Assuming a slow variation of the charge density inside a grid cell, a new Green function can be defined as:

$$G_i(r,r')=\int_{r'-\Delta/2}^{r'+\Delta/2}G(r,r')dr' \quad (5)$$

where $\Delta$ is the size of the grid cell. This integrated Green function can be used in the original FFT-based method to calculate the cyclic summation of the electric potential. The detailed expression of the integrated Green function in 2D can be found in Refs. [3–5] and in 3D can be found in Refs. [6–8].

The above method calculates the electric potential inside the beam itself. In some applications, one might also be interested in the fields outside the beam such as in the case of long-range beam–beam interaction or the image space charge forces from a flat conducting plate. Under these situations, the direct use of the above method will be inefficient since it requires using a computational domain that contains both the domain of the beam and the domain of the field that could be far from the beam. It is wasteful to define such a large computational domain since one is only interested in the fields inside the field domain. A more efficient method is to define a computational domain that is large enough to contain either the particle beam domain or the field domain and a new shifted Green function as [9]:

$$G_s(r,r')=G(r+r_s,r') \quad (6)$$

where $r_s$ is the separation distance between the particle beam domain and the field domain. This shifted Green function can be used in the above FFT-based method to calculate the cyclic summation.

The FFT-based methods above assume a uniform computational grid. In some applications, the particle density distribution is not uniform and a non-uniform grid might be preferred. For the 2D Poisson equation with open boundary condition, the solution of electric potential can be written as:

$$\phi(r,\theta)=\int G(r,r',\theta,\theta')\rho(r',\theta')r'dr'd\theta' \quad (7)$$

in a polar coordinate system, where:

$$G(r,r',\theta,\theta')=-\frac{1}{2}\log(r^2-2rr'\cos(\theta-\theta')+r'^2) .$$

The above convolution cannot be directly calculated using the FFT-based method. Instead, we define a new variable:

$$s = \frac{1}{k_1} \log\left(\frac{r}{k_2}\right) \quad (8)$$

Under this new variable, the Green function can be rewritten as

$$G(s,s',\theta,\theta') = -\frac{1}{2}\log(e^{2k_1(s-s')} - 2e^{k_1(s-s')}\cos(\theta-\theta') + 1)$$

The new Green function on the uniform grid of $s$ and $\theta$ can be calculated using the FFT-based method discussed above. This yields a non-uniform grid along the radial direction $r$. Figure 3 shows the electric field error calculated for a round beam with a Gaussian density distribution [10]. It is seen that in this case the non-uniform grid Green function method yields a factor of three less errors than the uniform grid Green function method.

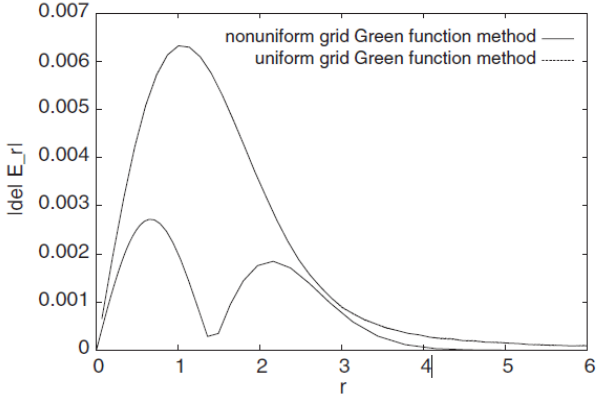

Figure 3: Electrical field error as a function of radius using the non-uniform grid Green method and the uniform grid Green function method.

## SPECTRAL FINITE DIFFERENCE METHOD

In some applications, the Poisson equation might be subject to simple regular shape boundary conditions. If an analytical eigenfunction can be found to satisfy both the Poisson equation and the boundary condition in a dimension, one can use the spectral method in that dimension and combine it with the finite difference method in the other dimensions. For example, a 2D Poisson equation subject to open boundary conditions can be written in cylindrical coordinates as:

$$\frac{1}{r}\frac{\partial}{\partial r}\left(r\frac{\partial}{\partial r}\phi\right) + \frac{1}{r^2}\left(\frac{\partial^2}{\partial \theta^2}\phi\right) = -\frac{\rho}{\varepsilon_0} \quad (9)$$

Making use of the periodic boundary condition in the azimuthal direction, the electric potential and the charge density can be written as:

$$\phi(r,\theta) = \sum \phi_m(r) e^{-im\theta}$$

$$\rho(r,\theta) = \sum \rho_m(r) e^{-im\theta}$$

Substituting those expressions into the Poisson's equation above yields:

$$\frac{1}{r}\frac{\partial}{\partial r}\left(r\frac{\partial}{\partial r}\phi_m\right) - \frac{m^2}{r^2}\phi_m = -\frac{\rho_m}{\varepsilon_0} \quad for \quad r \le a \quad (9a)$$

$$\frac{1}{r}\frac{\partial}{\partial r}\left(r\frac{\partial}{\partial r}\phi_m\right) - \frac{m^2}{r^2}\phi_m = 0 \quad for \quad r > a \quad (9b)$$

for each mode m. The ordinary differential equation for each mode inside the beam boundary can be solved using the finite difference method:

$$\left(\frac{1}{h^2} + \frac{1}{hr}\right)\phi_m^{n+1} - \left(\frac{2}{h^2} + \frac{m^2}{r^2}\right)\phi_m^n + \left(\frac{1}{h^2} - \frac{1}{hr}\right)\phi_m^{n-1} = -\frac{\rho_m}{\varepsilon_0} \quad (10)$$

subject to the boundary conditions:

$$\frac{\partial}{\partial r}\phi_m = 0 \quad for \quad r=0 \quad and \quad m=0$$

$$\phi_m = 0 \quad for \quad r=0 \quad and \quad m>0$$

For the ordinary differential equation outside the beam, the electric potential solution can be written as:

$$\phi = c\, r^{-m} \quad m > 0$$

$$\phi = c\, \ln(r) \quad m = 0$$

The above solutions can be used as a boundary condition to match the electric solution inside the beam. This solution together with the boundary condition at the origin will close the $N$ algebra equations for $N$ unknowns at each mode $m$. This group of linear algebra equations form a tri-diagonal equation that can be solved by direct elimination method with a computational cost of $O(N)$ where $N$ is the number of grid points inside the beam. Figure 4 shows the electric field as a function of radius for a round Gaussian beam from the numerical solution and the analytical solution. It is seen that the agreement between those two solutions is excellent.

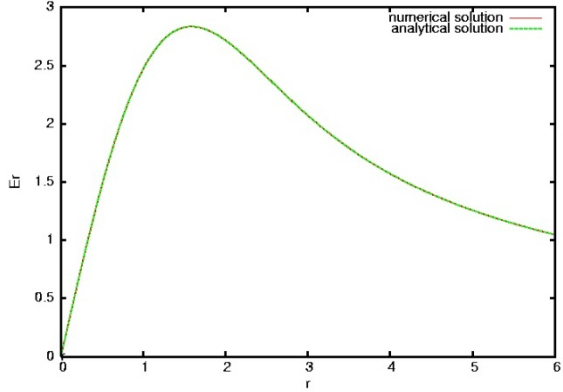

Figure 4: Electric field as a function of radius from the numerical solution and the analytical solution.

The above boundary matching method uses the computational domain to contain only the beam instead of the large empty space where the electric potential vanishes at the boundary. A similar numerical method was also used to solve the 3D Poisson equation subject to the transverse finite round and rectangular boundary conditions and the longitudinal open boundary condition [11].

## MULTIGRID SPECTRAL FINITE DIFFERENCE METHOD

In applications where the transverse boundary geometry is not regular, e.g. with an electrode, the simple spectral finite difference method above might not be applicable. In this case, a multigrid spectral method can be used. Given the 3D Poisson equation in a cylindrical coordinate system:

$$\frac{\partial^2 \phi}{\partial r^2} + \frac{1}{r}\frac{\partial \phi}{\partial r} + \frac{1}{r^2}\frac{\partial^2 \phi}{\partial \theta^2} + \frac{\partial^2 \phi}{\partial z^2} = -\frac{\rho}{\varepsilon} \quad (11)$$

subject to an azimuthally symmetric boundary condition, the charge density and electrical potential can be written as:

$$\rho(r,\theta,z) = \sum \rho^m(r,z)\exp(-im\theta)$$

$$\phi(r,\theta,z) = \sum \phi^m(r,z)\exp(-im\theta).$$

Substituting the above solutions into the Poisson equation yields:

$$\frac{\partial^2 \phi^m}{\partial r^2} + \frac{1}{r}\frac{\partial \phi^m}{\partial r} - \frac{m^2}{r^2}\phi^m + \frac{\partial^2 \phi^m}{\partial z^2} = -\frac{\rho^m}{\varepsilon}. \quad (12)$$

The above group of partial differential equations can be solved for each mode $m$ using a finite difference method with appropriate boundary geometry shape, which results in a group of algebraic equations. These algebraic equations form a sparse matrix equation and can be solved using an iterative method. Directly solving the above sparse matrix equation on the original grid using a classical iterative matrix-vector multiplication method such as the successive over-relaxation method (SOR) has a slow convergence rate. This is because the low-frequency errors during the iteration decrease slowly after the first few iterations. The classical iteration method moves the information one grid per iteration and will take a large number of iterations ($O(N^{1/d})$) (where $d$ is the dimension of the problem) to move the information across the full grid. Since the operation is done on the full grid, this results in a large number of operations to solve the linear algebraic equations.

The multigrid method smoothes out the numerical errors of different frequencies on different scales using multiple grids. It moves the information across the grid using $O(\log(N))$ steps. Most matrix-vector multiplications are done on the coarser grid with a much smaller number of operations so that the total number of operations in the multigrid method scales as $O(N)$. It replaces the error correction on the finer grid by an approximation on the coarser grid. It solves the coarser grid problem recursively by using a still coarser approximation until the direct solutions can be found on the coarsest grid. The solution from the coarser grid is then interpolated back to the finer grid and is used as a new starting solution on the finer grid until the final finest grid is reached. A detailed discussion of the multigrid method can be found in Refs. [12, 13] and an application to the simulation of the ion beam formation from an ECR ion source can be found in reference [14].

## REFERENCES

[1] R.W. Hockney and J.W. Eastwood, Computer Simulation Using Particles (New York: McGraw-Hill Book Company, 1985).
[2] W.H. Press et al., Numerical Recipes in FORTRAN: The Art of Scientific Computing, 2nd ed. (Cambridge: Cambridge University Press, 1992).
[3] K. Ohmi, Phys. Rev. E 62 (2000) 7287.
[4] R.D. Ryne, ICFA Beam Dynamics Mini Workshop on Space Charge Simulation, Trinity College, Oxford (2003).
[5] J. Qiang, M. Furman, and R. Ryne, J. Comp. Phys. 198 (2004) 278.
[6] J. Qiang et al., PRST-AB 9, (2006) 044204.
[7] J. Qiang et al., PRST-AB 10, (2007) 129901.
[8] V. Ivanov, Int. J. Mod. Phys. A, 24, (2009) 869.
[9] J. Qiang, M. Furman, and R. Ryne, PRST-AB 5, (2002) 104402.
[10] J. Qiang et al., NIM-A, 558, (2006) 351.
[11] J. Qiang and R. Ryne, Comput. Phys. Comm. 138 (2001) 18.

[12] W. Hackbusch, Multi-Grid Methods and Applications (New York: Springer-Verlag, 1985).
[13] P. Wesseling, An Introduction to Multigrid Methods (Chichester: John Wiley& Sons, 1992).
[14] J. Qiang, D. Todd, and D. Leitner, Comput. Phys. Comm. 175 (2006) 416.